\documentclass{webofc}
\usepackage[varg]{txfonts}   
\usepackage{graphicx}
\usepackage{here}

\usepackage{wrapfig}

\usepackage{stackrel}
\usepackage{amssymb,accents,latexsym}
\usepackage{color}

\newcommand{\be}{\begin{equation}}
\newcommand{\ee}{\end{equation}}
\newcommand{\bea}{\begin{eqnarray}}
\newcommand{\eea}{\end{eqnarray}}

\newcommand{\tr}{{\textrm {tr}}}

\newcommand{\h}{\mathfrak h}

\newcommand{\KK}{\textrm{\scriptsize KK}}
\newcommand{\eff}{\textrm{\scriptsize eff}}

\newcommand{\GHY}{\!\textrm{\tiny GHY}}

\newcommand{\Imaginary}{{\textrm{Im}}}

\newcommand{\EFT}{\textrm{\tiny EFT}}
\newcommand{\Pl}{\textrm{\scriptsize Pl}}

\begin{document}
\title{Physics of warped dimensions: discrete and continuous spectra~\thanks{Talk given by E.~Meg\'{\i}as at the XVth Quark Confinement and the Hadron Spectrum Conference, 1-6 August 2022, Stavanger, Norway.}}
%
%

\author{\firstname{Eugenio} \lastname{Meg\'{\i}as}\inst{1}\fnsep\thanks{\email{emegias@ugr.es}} \and
  \firstname{Mariano} \lastname{Quir\'os}\inst{2}\fnsep\thanks{\email{quiros@ifae.es}}
}

\institute{
  Departamento de F\'{\i}sica At\'omica, Molecular y Nuclear and Instituto Carlos I de F\'{\i}sica Te\'orica y Computacional, Universidad de Granada, Avenida de Fuente Nueva s/n, 18071 Granada, Spain
  \and
  Institut de F\'{\i}sica d'Altes Energies (IFAE), Barcelona Institute of  Science and Technology (BIST), Campus UAB, 08193 Bellaterra, Barcelona, Spain
          }

\abstract{%
  Using two different warped five-dimensional (5D) models with two branes along the extra dimension, we study the Green's functions and the spectral properties of some of the fields propagating in the bulk. While the first model has a discrete spectrum of Kaluza-Klein (KK) modes, the second one has a continuous spectrum above a mass gap. We also study the positivity of the spectral functions, as well as the coupling of the graviton and the radion with SM matter fields.
}
\maketitle

\section{Introduction}
\label{Sec:intro}

The Standard Model (SM) of particle physics has passed all the tests at past and present experiments (Tevatron, LHC, $\dots$), and no clear deviation has been found so far. However, the SM fails to describe a number of observational and theoretical aspects (dark matter, hierarchy problem, $\dots$), so that  it is believed that the SM is an effective theory that should be completed in the ultraviolet (UV). One of the most fruitful completions is provided by the Randall-Sundrum (RS) model~\cite{Randall:1999ee}, in which the hierarchy between the Planck and the electroweak (EW) scales is generated by a warped extra dimension in anti-de Sitter (AdS$_5$) space. This theory predicts a discrete spectrum of Kaluza-Klein (KK) states associated with each SM field, with masses $m_\KK \sim \textrm{TeV}$. The elusiveness of isolated and narrow resonances in direct new physics searches at colliders motivated to study different solutions, including the clockwork models~\cite{Giudice:2016yja} and the linear dilaton (LD) models~\cite{Antoniadis:2011qw,Cox:2012ee}. Partly motivated by these works, a new kind of warped models, characterized by the existence of a continuum of states heavier than a TeV mass gap, have been recently proposed~\cite{Csaki:2018kxb,Megias:2019vdb,Megias:2021arn}. These models have recently been shown to have implications not only for particle physics, but also for cosmology~\cite{Csaki:2021gfm,Fichet:2022ixi}. The physics behind models with a gapped continuum spectrum is related to unparticles and unhiggs theories~\cite{Georgi:2007ek,Falkowski:2008yr}.

In the present work we study the Green's functions of several kind of fields in two different warped models: i) the RS model, and ii) the LD model. Thus, we will be able to provide details on the spectral properties of theories with either discrete or continuous spectra.

\section{The extra-dimensional model}
\label{sec:Model}

The warped extra-dimensional models are based on a 5D space-time with line element
\begin{equation}
ds^2 = g_{MN} dx^M dx^N = \bar g_{\mu\nu} dx^\mu dx^\nu - dy^2 \,,
\end{equation}
where $y$ is the extra dimension, and $\bar g_{\mu\nu} = e^{-2A(y)} \eta_{\mu\nu}$ is the 4D induced metric. Let us consider a scalar-gravity system with two branes located at $y = y_0$ (UV brane) and $y = y_1$ (IR brane), where we are fixing $y_0 = 0$ and $A(y_0) = 0$. The 5D action of the model reads~\cite{Cabrer:2009we}
\begin{equation}
S = \int d^5x \sqrt{|g|} \left[ -\frac{1}{2\kappa^2} R + \frac{1}{2} g^{MN}(\partial_M \phi)(\partial_N \phi) - V(\phi) \right] -\sum_{\alpha} \int_{B_\alpha} d^4x \sqrt{|\bar g|} \lambda_\alpha(\phi) + S_{\GHY} \,,
\end{equation}
where $V(\phi)$ is the bulk scalar potential, $\lambda_\alpha(\phi)$ are the UV $(\alpha=0)$ and IR $(\alpha = 1)$ 4D brane potentials, and $\kappa^2 = 1/(2 M_5^3)$ with $M_5$ being the 5D Planck scale. The equations of motion (EoM) of the scalar field in the bulk can be written in terms of the superpotential as~\cite{DeWolfe:1999cp}
\begin{equation}
\phi^\prime(y) = \frac{1}{2} W^\prime(\phi) \,, \qquad A^\prime(y) = \frac{\kappa^2}{6} W(\phi) \,,
\end{equation}
where the prime stays for derivative with respect to the corresponding argument, while the brane potentials are responsible for boundary or jumping conditions of the fields in the branes. Simple brane potentials satisfying these conditions are given by $\lambda_\alpha(\phi) = (-1)^\alpha W(\phi) + \frac{1}{2} \gamma_\alpha (\phi - v_\alpha)^2$ for $\alpha = 0, 1$. Moreover, by using the EoM, one can express the on-shell action as the 4D integral $S = - \int d^4x U_{\eff}$, where the effective potential $U_{\eff}$ is~\footnote{We are providing in Sec.~\ref{sec:Model} formulas that are valid for the RS model with two branes (RS1), a model defined in the domain $y \in [y_0,y_1]$, cf. Sec.~\ref{sec:RS}. In the LD model of Sec.~\ref{sec:LDM}, the domain is larger: $y \in [y_0,y_s)$ with $y_1 < y_s$; and in this case the bulk contribution in~$U_{\eff}$ vanishes at the singularity, $y_s$, and one has instead $U_1(\phi) = \lambda_1(\phi) = \frac{1}{2} \gamma_1 (\phi - v_1)^2$.}
\begin{equation}
U_{\eff} = \left[ e^{-4A} U_0(\phi) \right]_{y_0} + \left[ e^{-4A} U_1(\phi) \right]_{y_1} \,, \label{eq:Ueff}
\end{equation}
with $U_\alpha(\phi) = \lambda_\alpha(\phi) - (-1)^\alpha W(\phi) = \frac{1}{2} \gamma_\alpha (\phi - v_\alpha)^2$. The potential $U_{\eff}$ fixes dynamically the brane distance and the values of $\phi$ at the branes, i.e.~$v_\alpha \equiv  \phi(y_\alpha)$. Solving the hierarchy problem demands that the brane dynamics fixes $A(\phi_1) - A(\phi_0) \approx 35$, so that $M_{\Pl} \simeq 10^{15} \cdot \rho$ where $M_{\Pl}$ is the 4D Planck scale and $\rho$ is a scale ${\mathcal O}(\textrm{TeV})$~\cite{Goldberger:1999wh}.

\section{Discrete spectra: Randall-Sundrum model}
\label{sec:RS}

The RS1 model is defined by the constant superpotential
\begin{equation}
W(\phi) = \frac{6k}{\kappa^2} \,,
\end{equation}
where $k = 1/\ell$, and $\ell$ is the radius of AdS$_5$. The solution of the EoMs gives $A(y) = k y$ and $\phi(y) = \textrm{cte}$ for $(y_0 \le y \le y_1)$, thus corresponding to the AdS$_5$ geometry.

\subsection{Gauge bosons}
\label{subsec:Gauge_bosons}

The 4D Lagrangian for massless gauge bosons, in the gauge $A_5 = 0$, writes
\begin{equation}
\mathcal L_{4D} = \int_{y_0}^{y_1} dy\left[ -\frac{1}{4} \tr \, F_{\mu\nu}F^{\mu\nu} + \frac{1}{2}e^{-2A} \, \tr \, \partial_y A_\mu \partial_y A^\mu  \right]\,, 
\end{equation}
where the gauge fluctuations can be written as $A_\mu(p,y) = f(y) A_\mu(p)/\sqrt{y_1}$. By using conformal coordinates, $ds^2 = e^{-2A(z)} ( \eta_{\mu\nu} dx^\mu dx^\nu - dz^2 )$ where $z = e^{ky}/k$, and defining a rescaled field by $f(z) = e^{A(z)/2}\tilde{f}(z)$, the EoM of the fluctuations is expressed in the Schr\"odinger-like form
\begin{equation}
  -\tilde{f}^{\prime\prime}(z) + V_A(z) \tilde{f}(z) = p^2 \tilde{f}(z)  \qquad \textrm{with} \qquad   V_A(z) = \frac{3}{4z^2} \,, \qquad (z_0 \le z \le z_1) \,,
\end{equation}
with $z_0 = 1/k$ and $z_1 = 1/\rho$, while $\rho = k \, e^{-A(z_1)}$. The fact that the model is defined in a compact domain, $z \in [z_0,z_1]$, leads to a discrete spectrum of KK modes. Then, the wave functions can be normalized to $\langle f_m | f_n \rangle = \int_{y_0}^{y_1} dy \,f_{m}^\ast(y) f_{n}(y) = y_1 \delta_{mn}$, and $\{|f_n\rangle\}$ forms an orthonormal basis of the space. We have used the Dirac notation:  $\langle y | f_{n} \rangle = f_{n}(y)$ and $\langle f_{n} | y \rangle = f_{n}^{\ast}(y)$.

\subsection{Green's functions for gauge bosons}
\label{subsec:GF_gauge_bosons}

The Green's functions for gauge bosons propagating in the bulk from $y$ to $y^\prime$ are given by $G^{\mu\nu}_A(y,y^\prime;p)=[\eta^{\mu\nu}-(1-\xi)p^\mu p^\nu/p^2]G_A(y,y^\prime;p)$, where $G_A$ obeys the following EoM
\begin{equation}
p^2 G_A(y,y^\prime;p) + \partial_y \left( e^{-2A(y)} \partial_y G_A(y,y^\prime;p) \right) = \delta(y-y^\prime) \,.  \label{eq:GAy}
\end{equation}
We are using a mixed representation where only the 4D coordinates $x^\mu$ are Fourier transformed into 4D momenta $p^\mu$. This equation can be solved analytically by imposing Neumann boundary conditions in the UV and IR branes,  i.e.~$(\partial_y G_A)(y_0)  = 0$ and $(\partial_y G_A)(y_1) = 0$, as well as the matching conditions $\Delta G_A(y^\prime) = 0$ and $\Delta(\partial_y G_A)(y^\prime) = e^{2A(y^\prime)}$. In these expressions $G_A(y)$ means $G_A(y,y^\prime)$ and $\Delta F(y) \equiv \lim_{\epsilon \to 0}\left(F(y+\epsilon) - F(y-\epsilon) \right)$. The result is given by
\begin{eqnarray}
G_{A}(y,y^\prime;p) = \frac{\pi}{2k} e^{k (y + y^\prime)}  \left[ Y_0\left( \frac{p}{k} \right)  J_1\left( e^{ky_\downarrow} \frac{p}{k}  \right)   - J_0\left( \frac{p}{k} \right)  Y_1 \left( e^{ky_\downarrow} \frac{p}{k} \right)   \right] \times \frac{Z_1(p,y_\uparrow)}{\Phi_A(p)}  \,,\label{eq:GAyyp}
\end{eqnarray}
with $Z_\alpha(p,y) = J_0\left( p/\rho \right)   Y_\alpha\left( e^{k y} p/k \right) -  J_\alpha\left( e^{k y} p/k \right) Y_0\left( p/\rho \right)$ and $\Phi_A(p) \equiv Z_0(p,y_0)$, while $J_n(z)$ and $Y_n(z)$ are the Bessel functions of the first and second kind, respectively. We have defined the variables $y_\downarrow =\textrm{min}(y,y^\prime)$ and $y_\uparrow=\textrm{max}(y,y^\prime)$. The Green's functions have poles at the zeros of $\Phi_A(p)$, and these correspond to the discrete KK mass spectrum.  The lowest-lying states of the spectrum are at $p_n \equiv m_n$ with $m_n/\rho \simeq 0,\;  2.45, \; 5.57, \; 8.70, \; 11.84, \; 14.98, \; \cdots$. Notice that there is a zero mode and an infinite tower of massive modes.

\subsection{Spectral properties of gauge bosons}
\label{subsec:spectral_functions}

\begin{figure*}[t]
  \centering
  \sidecaption
  \includegraphics[width=4.5cm]{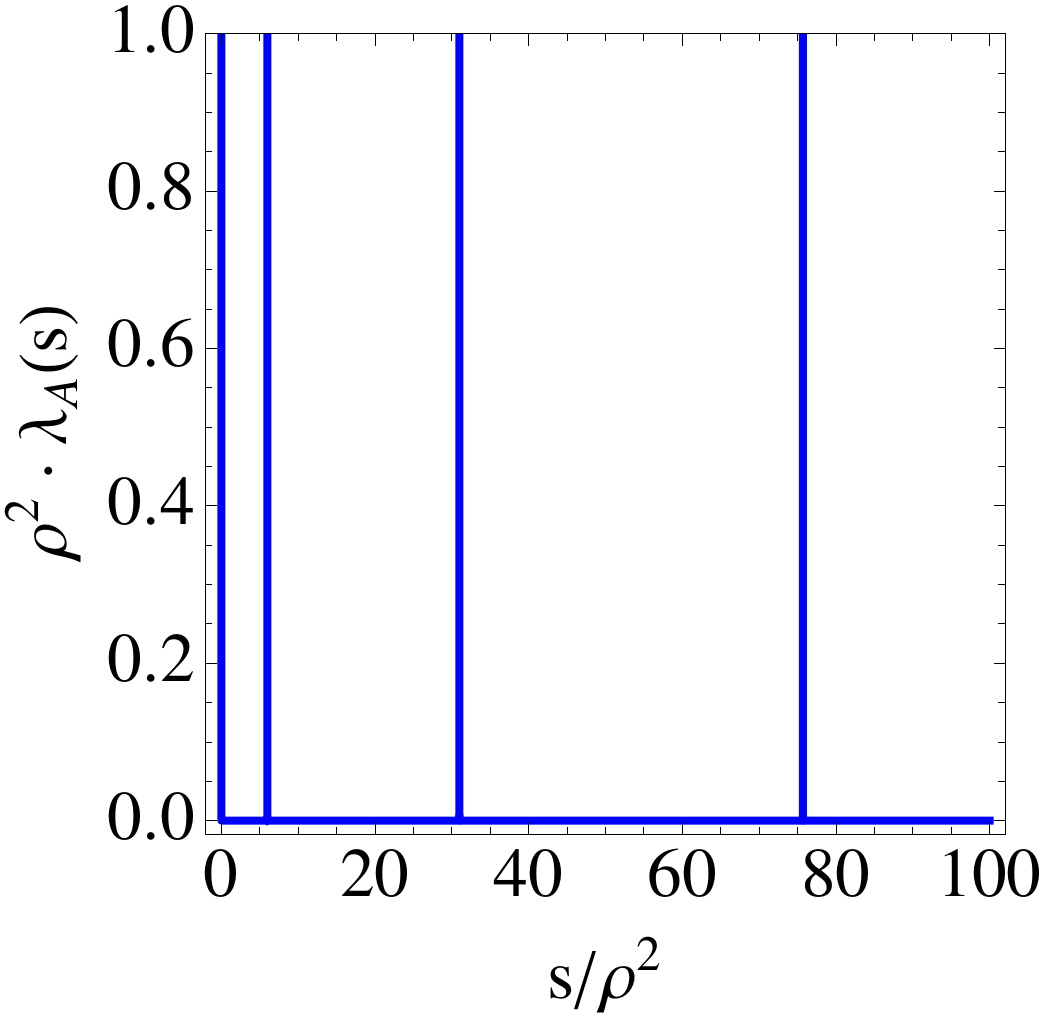} 
 \caption{Density of states $\lambda_A$ (normalized by $\rho^2$), as a function of $s/\rho^2$, for the KK modes of the gauge bosons in the RS model, cf. Eq.~(\ref{eq:lambda_RS}). We have used $A_1 = 35$.}
\label{fig:Density_RS}
\end{figure*}

The spectral functions are defined as
\begin{equation}
\rho_A(y,y^\prime;s) = -\frac{1}{\pi} \Imaginary \, G_A(y,y^\prime; s + i 0^+) \,, \qquad s \equiv p^2 \,. \label{eq:rhoA}
\end{equation}
It would be clarifying to study these functions in an eigenvalue decomposition. The Green's functions and spectral functions can be understood as matrix elements of the operators~\footnote{The operators of Eq.~(\ref{eq:rho_RS}) are related by $\hat \rho_{A}(s) = - \frac{1}{\pi} \Imaginary \; \hat G_{A}(s)$, where $\Imaginary \; \hat G_{A}(s) = \frac{1}{2i} \left( \hat G_{A}(s) - \hat G_{A}^\dagger(s) \right)$.
}
\begin{equation}
\hat G_{A}(s) = \frac{1}{y_1} \sum_n \frac{ | f_{n} \rangle \langle f_{n}| }{s - m_n^2 + i 0^+} \,,  \qquad \hat\rho_{A}(s) = \frac{1}{y_1} \sum_n | f_{n} \rangle \langle f_{n} |  \; \delta(s - m_n^2)\,,  \label{eq:rho_RS}
\end{equation}
where $f_n(y)$ are the wave functions defined in Sec.~\ref{subsec:Gauge_bosons}, i.e. $G_{A} (y,y^\prime;s) = \langle y | \hat G_{A}(s) |y^\prime \rangle$ and $\rho_{A}(y,y^\prime;s) = \langle y | \hat \rho_{A}(s) |y^\prime \rangle$. Then, by using the Dirac notation of quantum mechanics, one has
\begin{equation}
G_{A}(y,y^\prime;s) = \frac{1}{y_1} \sum_n \frac{f_{n}(y) f_{n}^{\ast}(y^\prime)}{s - m_n^2 + i 0^+} \,, \qquad \rho_{A}(y,y^\prime;s) = \frac{1}{y_1} \sum_n f_{n}(y) f_{n}^{\ast}(y^\prime)  \; \delta(s - m_n^2) \,. \label{eq:rhomn_RS}
\end{equation}
From these two expressions one obtains the spectral representation of the Green's function $G_{A}(y,y^\prime;s) = \int dm^2 \frac{\rho_{A}(y,y^\prime;m^2)}{s-m^2 +  i 0^+}$. One can see from Eq.~(\ref{eq:rhomn_RS}) that while $\rho_A(y,y;s)$ is positive definite, $\rho_A(y,y^\prime;s)$ with $y \ne y^\prime$ is not, as $f_{n}(y)$ has $n$ nodes and so it has $n$ changes of sign in $(y_0, y_1)$.  This apparent contradiction challenges the physical interpretation of the spectral functions in 4D QFT. Let's point out that what should be positive semidefinite is the operator $\hat \rho_A(s)$, so that while this implies that all its eigenvalues are positive semidefinite, this does not mean that all its matrix elements $\langle y | \hat \rho_A(s) | y^\prime \rangle$ are. This will be briefly addressed now in the context of the RS model, and more generically in Sec.~\ref{subsec:gravition} within a model with continuous spectrum. By using Eq.~(\ref{eq:rho_RS}), one gets that $|f_n\rangle$ is an eigenvector of the spectral operator $\hat\rho_{A}(s)$ with eigenvalue $\delta(s-m_n^2)$, i.e. $\hat\rho_{A}(s) | f_n\rangle = \delta(s - m_n^2) | f_n\rangle$. Then, we find that in the off-shell case $(s \ne m_n^2$ $\forall n)$ all the eigenvalues of $\hat\rho_{A}(s)$ are vanishing, while in the on-shell case $(s = m_n^2)$ there is a single non-vanishing eigenvalue which is infinite~$\delta(0)$, and the others are vanishing. Moreover, the trace of the spectral operator, which is the summation of all its eigenvalues,
\begin{equation}
\lambda_{A}(s) \equiv \tr \, \hat\rho_{A}(s) = \sum_n \delta(s - m_n^2) \ge 0 \,, \label{eq:lambda_RS}
\end{equation}
can be interpreted as the {\it density of states} in momentum $s$. Finally, the integral
\begin{equation}
\int_0^\infty ds \, \lambda_{A}(s) = \int_0^\infty ds \sum_n \delta(s - m_n^2)  =  N_{\textrm{states}} \to \infty 
\end{equation}
is the {\it number of states}, a quantity that turns out to be divergent as there are infinite states in the spectrum. We display in Fig.~\ref{fig:Density_RS} the density of states in the RS model as a function of $s/\rho^2$. Notice that each KK mode contributes with a Dirac delta function with $\int_{m_n^2 - \epsilon}^{m_n^2 + \epsilon} ds \, \lambda_{A}(s) = 1$. 

\section{Gapped continuous spectra: linear dilaton model}
\label{sec:LDM}

When new physics consists in heavy Breit-Wigner resonances, their presence is detected by bumps in the invariant mass of the final states, corresponding to the mass of the exchanged particle. The lack of detection of bumps associated with states not present in the SM motivated  the study of models with a continuum of KK states beyond a mass gap, $m_g$. In this scenario, new physics is associated with an excess in the measured cross section with respect to the SM prediction~\cite{Csaki:2018kxb,Megias:2019vdb}. The Green's functions that we will study in this section generalize the particle propagators with isolated poles $(p^2 - m^2 + i 0^+)^{-1} = \mathcal{P} (p^2 - m^2)^{-1} - i \pi \delta(p^2 - m^2)$, to Green's functions with an isolated pole (zero mode) and a continuum of states above $m_g$,
\begin{equation}
G(p^2,m_g^2) = \textrm{Re} \, G(p^2,m_g^2) + i \, \left[ c_0 \delta(p^2) + \eta(p^2,m_g^2) \Theta(p^2 - m_g^2) \right] \,.
\end{equation}
This is the behavior of gapped unparticles~\cite{Falkowski:2008yr,Falkowski:2009uy}. In the extra-dimensional model, the gap $m_g \sim \rho \sim$ TeV is linked to the solution of the hierarchy problem as discussed in Sec.~\ref{sec:Model}.

\subsection{The linear dilaton model}
\label{subsec:LDM}

The LD model is defined by the superpotential~\cite{Megias:2021mgj}
\begin{equation}
W(\bar\phi) = \frac{6k}{\kappa^2} e^{\bar\phi}  \,,
\end{equation}
where  $\bar\phi \equiv \kappa \phi / \sqrt{3}$. The warped factor and scalar field have a linear behavior in conformal coordinates, i.e. $A(z) = \rho \cdot (z - z_0)$ and $\bar\phi(z) = A(z) + \bar v_0$ for $(z_0 \le z < \infty)$, with $z_0 = 1/k$ and $\rho = \pm 1/y_s$.~\footnote{In proper coordinates, the solution is $A(y) = - \log\left(1 - y/y_s \right)$ and $\bar \phi(y)  = - \log\left[k (y_s - y) \right]$ for $(y_0 \le y < y_s)$, where $y = y_s$ is an admissible singularity in the IR.} This model leads to continuous or discrete spectra (depending on the sign of~$\rho$) for all the fields, with gaps $m_g \sim \rho$. We can consider two situations:
\begin{itemize}
\item The case $\rho = - 1/y_s$ is dual to Little String Theories~\cite{Antoniadis:2011qw,Cox:2012ee}. Using $\kappa^2 M_{\textrm{Pl}}^2 = \int_{z_0}^{z_1} dz  \, e^{-3A}$, the relationship between the 5D and 4D Planck scales is $M_5 \simeq \left( \frac{3}{2}  |\rho| M_{\textrm{Pl}}^2 \right)^{1/3} e^{-|A_1|} \sim |\rho| \sim \textrm{TeV}$, while $A_1 \equiv A(z_1) \simeq 23$. In this theory the hierarchy problem has to be entirely solved by a string theory which should set the cutoff scale at the TeV. In this case no continuum spectrum is allowed,  as an infinite value of $z_1$ would imply $M_{\textrm{Pl}} \to \infty$, so that gravity would be decoupled. Then, the spectrum is discrete and it has a gap, $m_n^2 =  m_g^2 +  \frac{\pi^2 n^2}{A_1^2} \rho^2 \,, \; (n \in \mathbb{Z})$.

\item The model for $\rho = +1/y_s$ can be defined in $z \in [z_0,+\infty)$. This leads to $M_{5} \simeq \left( \frac{3}{2} \rho M_{\textrm{Pl}}^2 \right)^{1/3} \simeq 10^{10} \cdot \rho$ for $\rho \sim \textrm{TeV}$ and $A_1 \simeq 23$. This theory solves the hierarchy problem between $M_5$ and the TeV scale, and a UV completion of the theory given by a string theory that would solve the hierarchy problem between $M_{\textrm{Pl}}$  and $M_5$ would be needed. This case leads to a  gapped continuous spectrum, $m^2 \ge m_g^2$.
\end{itemize}

In the following we will assume $\rho = +1/y_s$. Within the LD model, propagation of gauge bosons in the bulk produces a violation of EW precision observables, so that in the following we will consider that: i) SM fields (gauge bosons, fermions and Higgs boson) are located in the IR brane $(z = z_1)$, and ii) the {\it graviton} and the {\it radion} can propagate in the bulk~\cite{Megias:2020cpw,Megias:2021mgj}.

\subsection{Graviton}
\label{subsec:gravition}

\begin{figure}[t]
\centering
\includegraphics[width=3.9cm]{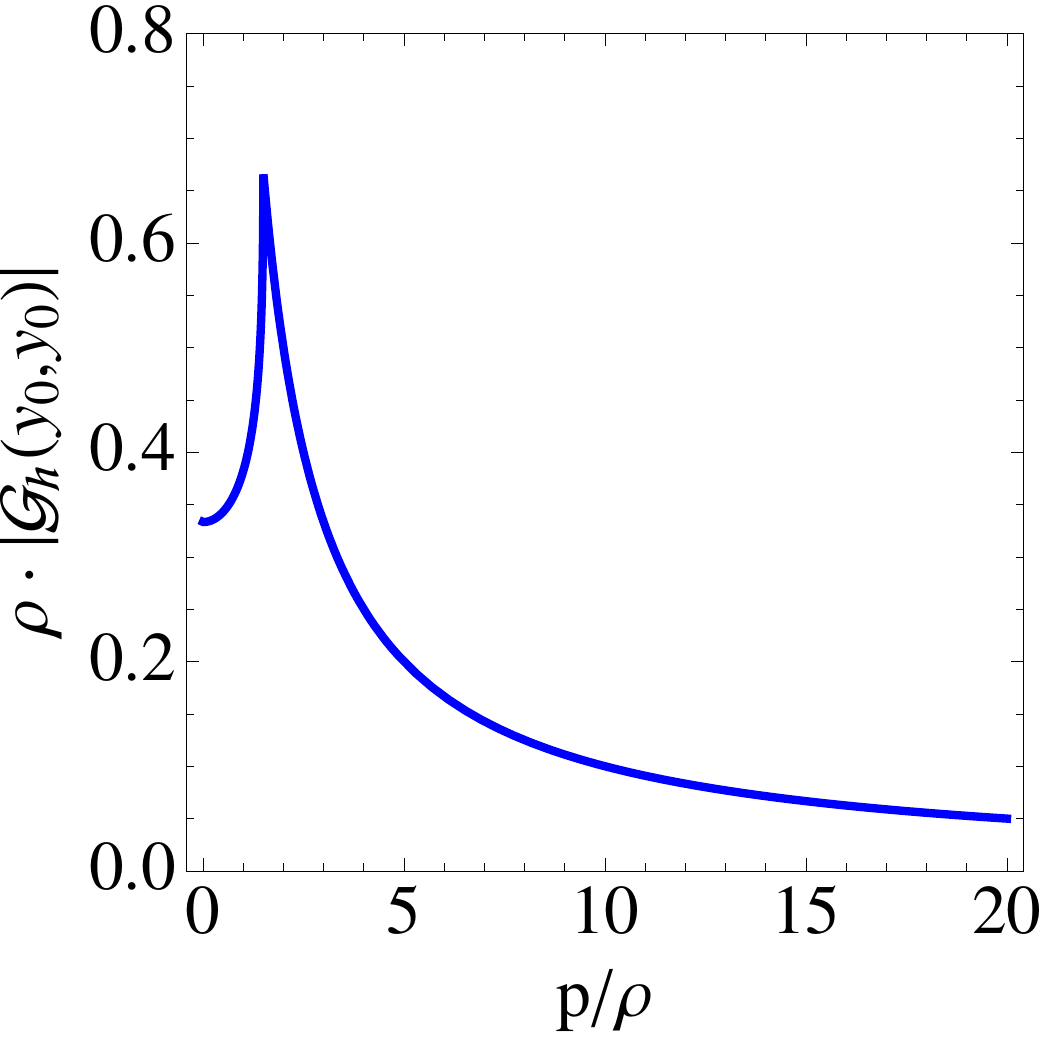} \hspace{0.3cm}
\includegraphics[width=4cm]{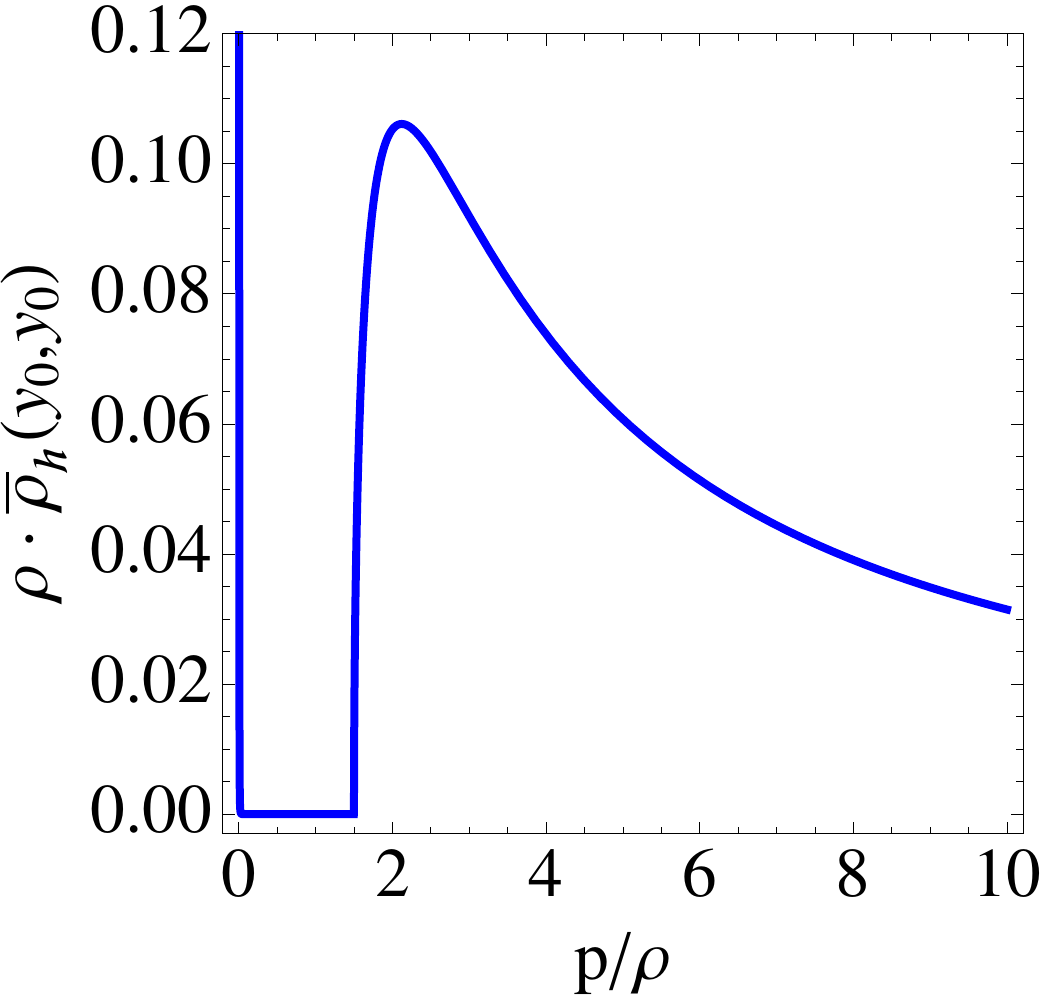} \hspace{0.3cm}
\includegraphics[width=3.9cm]{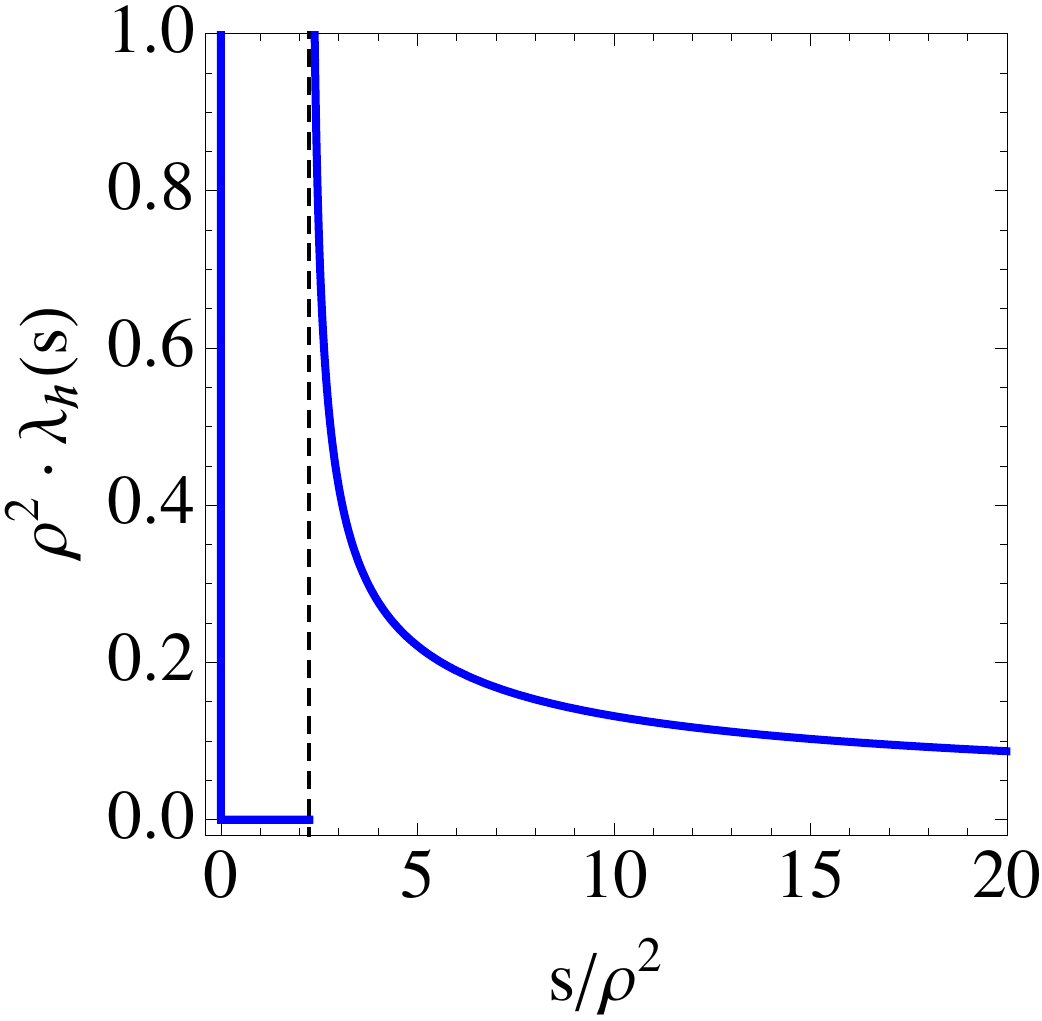}
\caption{\it Plots of the Green's function $|\mathcal G_\h(y_0,y_0;p)|$ (left panel), spectral function $\bar\rho_\h(y_0,y_0;p)$ (middle panel), and density of states $\lambda_\h(s)$ (right panel), as a function of the momentum, for a continuous graviton, cf. Eqs.~(\ref{eq:Gh_yyp}) and (\ref{eq:lambda_LD}).  The dashed vertical line in the right panel corresponds to $s = m_g^2$. We have used $\bar\epsilon = 0.1$, $A_1 = 23$ and assumed time-like momenta $p^2>0$.}
\label{fig:Ggraviton}
\end{figure}

The graviton is a transverse traceless fluctuation of the metric of the form~\cite{Cabrer:2009we}
\begin{equation}
ds^2 = e^{-2A(y)} (\eta_{\mu\nu} + 2\kappa h_{\mu\nu}(x,y) ) dx^\mu dx^\nu - dy^2 \,.
\end{equation}
By using the ansatz $h_{\mu\nu}(x,y) = \h(y) h_{\mu\nu}(x)$, one can obtain the EoM of the graviton fluctuation, which in conformal coordinates and after defining the rescaled graviton fluctuation by $\h(z) = e^{3A(z)/2} \tilde\h(z)$, turns out to be $-\tilde\h^{\prime\prime}(z) + V_{\h}(z) \tilde{\h}(z) = p^2 \tilde{\h}(z)$ with $V_\h(z)  = (3\rho/2)^2$. The value $m_g = 3\rho/2$ corresponds to the mass gap in the spectrum. The Green's function for $h_{\mu\nu}(x,y)$ in the transverse, traceless gauge, obeys the EoM
\begin{equation}
p^2 G_\h(y,y^\prime;p) + e^{2A(y)} \partial_y \left( e^{-4A(y)} \partial_y G_\h(y,y^\prime;p) \right) = e^{2A(y^\prime)} \cdot \delta(y-y^\prime) \,,  \label{eq:Ghy}
\end{equation}
for $y_0 \le y,\, y^\prime < y_s$. The boundary and matching conditions are $(\partial_y G_\h)(y_0) = 0$ and $\Delta (\partial_y G_\h)(y^\prime) = e^{4A(y^\prime)}$, where $G_\h(y) \equiv G_\h(y,y^\prime)$. We should also impose continuity of $G_\h(y)$ in $y = y^\prime, y_1$, and $(\partial_y G_\h)(y)$ in $y = y_1$, as well as regularity in the IR $(y\to y_s)$. The result is~\cite{Megias:2021mgj}
\begin{equation}
G_\h(y,y^\prime;p) =  \frac{(1 - \bar y_\uparrow)^{\frac{3}{2}\Delta^+(p)}}{3\rho\delta(p)}\left(  - (1 - \bar y_\downarrow)^{\frac{3}{2}\Delta^-(p)} + \frac{\Delta^-(p)}{\Delta^+(p)} (1 - \bar y_\downarrow)^{\frac{3}{2} \Delta^+(p)}  \right) \,, \label{eq:Gh_yyp}
\end{equation}
where $\Delta^\pm(p) = \pm \delta(p) - 1$ and $\delta(p) = \sqrt{1 - (4/9)\cdot p^2/\rho^2}$, and we have defined $\bar y = \rho y$.  We display in Fig.~\ref{fig:Ggraviton} (left) the absolute value of the UV-UV Green's function with the zero mode subtracted out, i.e. ${\mathcal G}_\h(y_0,y_0) \equiv G_\h(y_0,y_0) - G_\h^0$ with $G_\h^0 = \frac{3\rho}{p^2} = \lim_{p\to 0} G_\h(y,y^\prime;p)$.

The spectral function in the LD model, defined as in Eq.~(\ref{eq:rhoA}), is related to a discontinuity of the Green's function at the branch cut $s \in [m_g^2,+\infty)$. Working in the basis with flat extra-dimensional coordinate $y$, i.e. $\bar h_{\mu\nu}(x,y) = e^{-A(y)} h_{\mu\nu}(x,y)$, an explicit evaluation leads to $\bar \rho_\h(y,y^\prime;s) = 3\rho e^{-A(y) - A(y^\prime)} \, \delta(s) + \bar \eta_\h(y,y^\prime;s) \Theta(s - m_g^2)$, where the terms $\propto \delta(s)$ and $\propto \bar\eta_\h(y,y^\prime;s)$ are the contributions from the zero mode and the continuum, respectively.  It can be proved that the operator $\hat\rho_\h(s)$, with matrix elements $(\hat\rho_\h)_{y y^\prime} \equiv \langle y | \hat \rho_\h (s)| y^\prime \rangle= \bar\rho_\h(y,y^\prime;s)$, is positive semidefinite, something discussed in Sec.~\ref{subsec:spectral_functions} for gauge bosons within the RS model.  To see that, let us point out that the infinite dimensional matrix $\hat\rho_\h$ has a factorizable form $(\hat \rho_\h)_{y y^\prime} = \rho_y \rho_{y^\prime}$ where $\rho_y = \sqrt{(\hat\rho_\h)_{yy}}$. Then, $\hat\rho_\h$ has the following properties: i) $\det \hat\rho_\h = 0$, and ii) all the eigenvalues of~$\hat\rho_\h$ are vanishing, except one of them, which is given by its trace,
\begin{equation}
\lambda_\h(s) = \tr \, \hat\rho_\h(s) = \int_0^{y_s} dy  \, \bar \rho_\h(y,y;s)  = \int_0^{y_s} dy \, \rho_y^2  \ge 0  \,. \label{eq:lambda_s}
\end{equation}
This finishes the proof. The integral of Eq.~(\ref{eq:lambda_s}) is divergent at $y = y_s$, and it can be regularized by performing the integral up to $y = y_s \cdot (1 - \bar\epsilon)$, where $\bar\epsilon$ is a regulator. The result is~\cite{Megias:2021mgj}
\begin{equation}
\lambda_\h(p) = \delta(p^2) + \left[ - \frac{\log \bar \epsilon}{2\pi \rho} \lambda_{\textrm{un}}(s) + {\mathcal O}(\bar\epsilon^0)\right] \,, \qquad \lambda_{\textrm{un}}(s) = (s - m_g^2)^{-1/2} \,, \label{eq:lambda_LD}
\end{equation}
where $\lambda_{\textrm{un}}(s)$ corresponds to an unparticle contribution with dimension $d_{\textrm{un}} = 3/2$~\cite{Delgado:2008gj}.

Within the LD model, we are assuming that the SM matter fields are located in the IR brane $y = y_1$. Then, the interaction Lagrangian in the 4D effective theory is given by
\begin{equation}
\mathcal L_{5D} = -\frac{1}{\sqrt{2}M_5^{3/2}} h^{\mu\nu}(x,y) T_{\mu\nu}(x,y) \delta(y-y_1) \,,
\end{equation}
where $T_{\mu\nu}(x,y_1)$ is the energy-momentum tensor (EMT) of matter. The zero mode, $h^0_{\mu\nu}(x,y)= h_0(y)h^0_{\mu\nu}$, couples as $-\frac{1}{M_{\rm Pl}} T^{\mu\nu}(x,y_1)h_{\mu\nu}^0(x)$. The coupling of the continuum KK modes with matter leads to an effective field theory (EFT) for $p \ll \rho$, which is $\mathcal L_{\EFT}(y_1) = c_\h(y_1) \mathcal O_\h(x,y_1)$, where $\mathcal O_\h(x,y_1) = T^{\mu}_{\ \nu} T_{\ \mu}^{\nu} -\frac{1}{3} (T^{\mu}_{\ \mu})^2$ and $c_\h(y_1)=  - g_{\rm eff}^2(y_1)/\rho^2$ with $g_{\rm eff}(y_1) = 1/(\sqrt{6}\rho)$.

\subsection{Radion}
\label{sec:radion}

\begin{figure}[t]
\centering
\includegraphics[width=3.9cm]{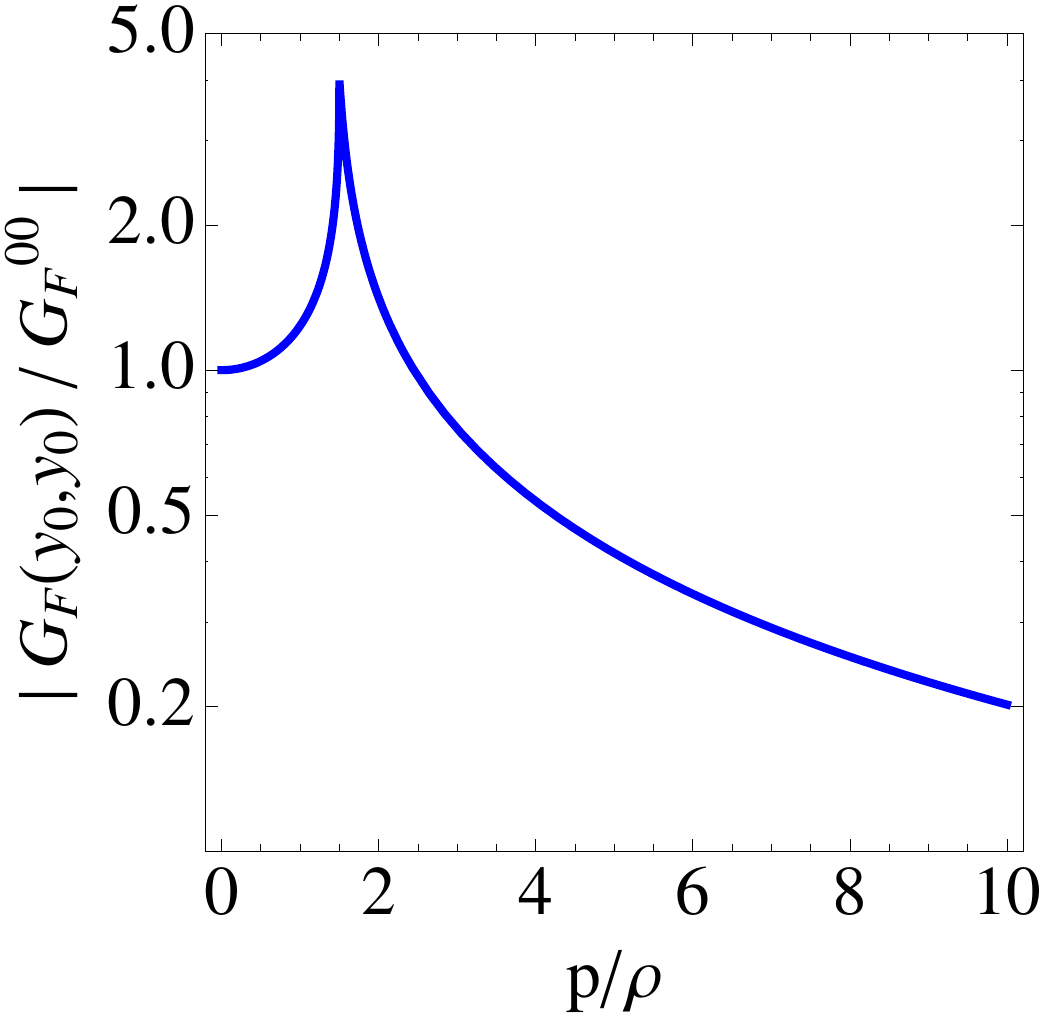} \hspace{0.3cm}
\includegraphics[width=4.2cm]{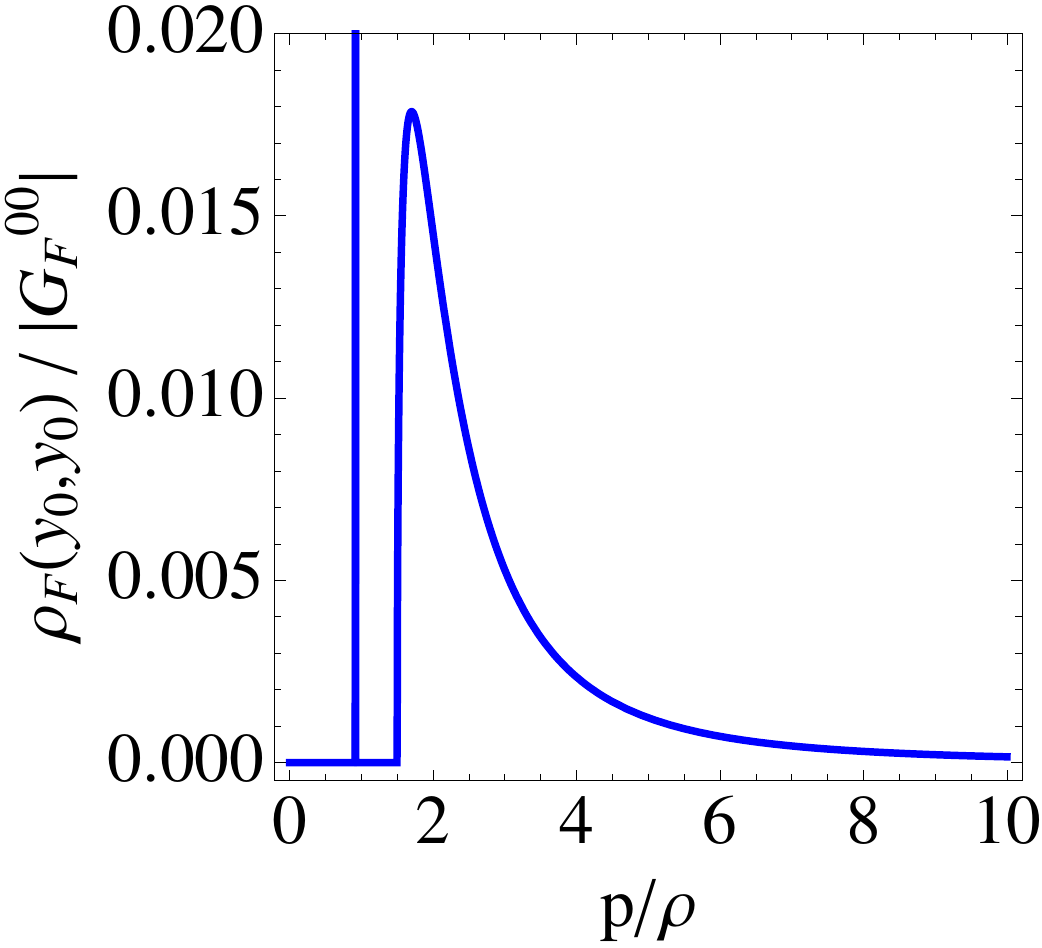} \hspace{0.3cm}
\includegraphics[width=3.8cm]{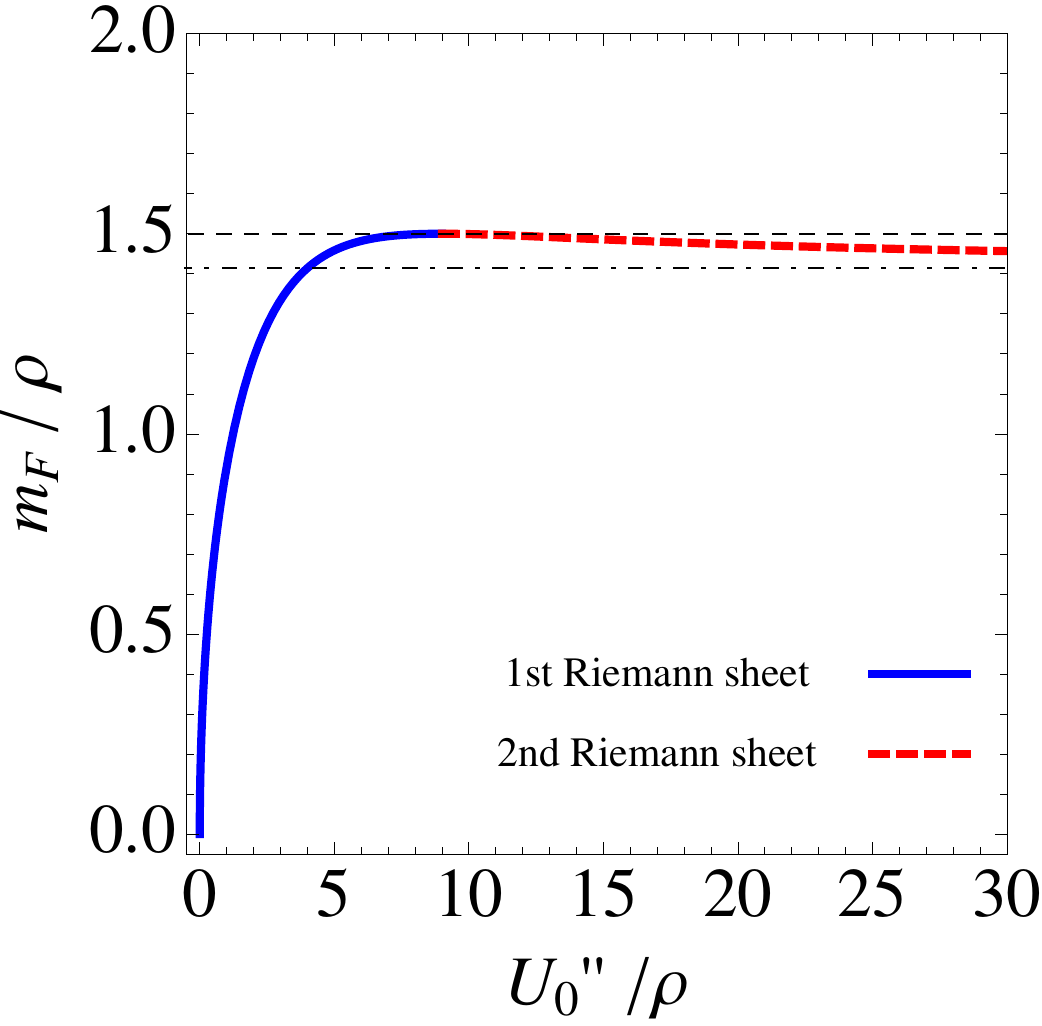} 
\caption{\it Plots of the Green's function $|G_F(y_0,y_0;p)|$  (left panel), spectral function $\rho_F(y_0,y_0;p)$ (middle panel), and dependence of the radion mass with $U_0^{\prime\prime}$, in the LD model. To guide the eye, we display in the right panel horizontal lines for values $m_F = \sqrt{2}\rho$ (dot-dashed black) and $m_F = m_g$ (dashed black). In the first two panels we have normalized the functions by $G_F^{00} \equiv \lim_{p\to 0} G_F(y_0,y_0;p) = -1/(2\rho)$. We have used $A_1 = 23$, and $U_0^{\prime\prime} = k$ (left panel) and $U_0^{\prime\prime} = \rho$ (middle panel).}
\label{fig:Radion}
\end{figure}

The radion field $F(x,y)$ is defined as the scalar perturbation of the metric~\cite{Csaki:2000zn}
\begin{equation}
ds^2 =  e^{-2(A(y) + F(x,y))} \eta_{\mu\nu} dx^\mu dx^\nu  - (1 + 2 F(x,y))^2 dy^2  \,,  \qquad \phi(x,y) = \phi(y) + \varphi(x,y) \,,
\end{equation}
with $F(x,y) = F(y) \mathcal R(x)$. The effective potential for the fluctuations is $V_F(z) = (3\rho/2)^2$, so that the radion spectrum has the same mass gap as the graviton. The EoM and boundary/mat\-ching conditions of the radion Green's function are the same as for the graviton, cf. Eq.~(\ref{eq:Ghy}), except for the boundary condition in the UV brane which is affected by the brane potential,
\begin{eqnarray}
  && (\partial_y G_F)(y_0) = \left( \frac{1}{3}\kappa^2 W(\phi(y)) - \frac{2p^2 e^{2A(y)}}{U_0^{\prime\prime}(\phi(y))} \right) G_F(y) \Bigg|_{y=y_0} \,,
\end{eqnarray}
where $G_F(y) \equiv G_F(y,y^\prime)$. Finally, the Green's function for the radion turns out to be~\cite{Megias:2021mgj}
\begin{align}
G_F(y,y^\prime;p) &= \frac{(1- \bar y_\uparrow)^{\frac{3}{2}\Delta^+(p)}}{3\rho\delta(p)}   \left[ - (1 - \bar y_\downarrow)^{\frac{3}{2}\Delta^-(p)}  +   \left( 1 +\frac{3 U_0^{\prime\prime}}{2\rho} \frac{\delta(p)}{\Phi_F(p)} \right)  (1-\bar y_\downarrow)^{\frac{3}{2}\Delta^+(p)} \right] \,, \label{eq:GF}
\end{align}
where $\Phi_F(p) = p^2/\rho^2 - (U_0^{\prime\prime}/(4\rho)) \cdot (1 + 3 \delta(p))$. The results for the UV-UV Green's function and spectral function for the radion are displayed in the left and middle panel of Fig.~\ref{fig:Radion}, respectively. Unlike the graviton case, $G_F(y,y^\prime;p)$ does not have an isolated massless mode, but a single massive mode corresponding to the zero of the function $\Phi_F(p)$, which is identified with the radion. We display in the right panel of Fig.~\ref{fig:Radion} the dependence of the radion mass, $m_F$, with the parameter~$U_0^{\prime\prime}$. Notice that the radion appears as a Dirac delta contribution in the spectral function. 
The radion couples to the SM fields through the trace of the EMT, i.e.
\begin{equation}
\mathcal L_{5D}=-\frac{1}{\sqrt{2} M_5^{3/2}}F(x,y)\mathcal T \delta(y-y_1) \,, \qquad  \mathcal T\equiv\tr \, T^\mu_{\ \nu}(x,y_1)\,.
\end{equation}
This gives rise, upon integration of the radion continuum, to the EFT Lagrangian with dimension eight operators $\mathcal L_{\EFT}=c_F(y_1) \mathcal T^2$, with Wilson coefficient $c_F(y_1) = c_\h(y_1)$.

\section{Conclusions}
\label{sec:Conclusions}

We have studied 5D warped models with two different geometries: i) AdS$_5$ and ii) linear dilaton. These models generate a hierarchy between the Planck scale and the EW scale by means of a warp factor, and a mechanism of dynamical stabilization of two branes, known as the Goldberger-Wise mechanism which, in the case of the RS model, is supposed not to appreciably backreact on the 5D metric. The KK spectra of all the particles are either: i)~discrete, or ii) continuous with a mass gap; depending on the domain of the theory in conformally flat coordinates: i) compact and ii) non-compact, respectively. We have computed in these models the Green's functions and spectral functions for the relevant fields: i) massless gauge bosons, and ii) graviton and radion; and identified the isolated poles of the Green's functions  with the discrete modes in the spectrum. We have also studied the positivity of the spectral functions, as well as the couplings of the continuum KK modes with the SM matter fields.

This work can be extended to the computation of the Green's functions of other fields (fermions, $\dots$), and to the analysis of the phenomenological implications of the existence of continuum spectra in the searches of new physics at colliders, as these will eventually increase the cross sections  $\sigma(pp\to Q\bar Q)$ with respect to the SM prediction~\cite{Csaki:2018kxb,Megias:2019vdb}. These applications, some of them inspired on unparticle phenomenology, will be addressed in future works.

\begin{acknowledgement}
\section*{Acknowledgement} 
We would like to thank S. Fichet and M. P\'erez-Victoria for
collaboration on related topics and enlightening discussions, and
L.L. Salcedo for discussions. The work of EM is supported by the
project PID2020-114767GB-I00 funded by MCIN/AEI/10.13039/501100011033,
by the FEDER/Junta de Andaluc\'{\i}a-Consejer\'{\i}a de Econom\'{\i}a
y Conocimiento 2014-2020 Operational Program under Grant
A-FQM-178-UGR18, and by Junta de Andaluc\'{\i}a under Grant
FQM-225. The research of EM is also supported by the Ram\'on y Cajal
Program of the Spanish MCIN under Grant RYC-2016-20678. The work of MQ
is partly supported by the Spanish MCIN under Grant FPA2017-88915-P,
by the Catalan Government under Grant 2017SGR1069, and by Severo Ochoa
Excellence Program of MCIN under Grant SEV-2016-0588. IFAE is
partially funded by the CERCA program of the Generalitat de Catalunya.
\end{acknowledgement}

%


%
%
%

\end{document}